\documentstyle[preprint,aps]{revtex}
\begin{document}
\draft
\title{The level spacing distribution near the Anderson transition\thanks{To
be published in JETP Letters, vol.\ 50, no 1 (1994)}
}
\author{ A.\  G.\  Aronov$^{1,2}$, V.\  E.\  Kravtsov$^{1,3}$, I.\  V.\
Lerner$^4$ }
\address{$^1$International Centre for Theoretical Physics,
P.O.\  Box 586, 34100 Trieste, Italy\\
$^2$A.\ F.\ Ioffe Physico-Technical Institute,
194021 St.\ Petersburg, Russia\\
$^3$Institute of Spectroscopy, Russian
 Academy of Sciences, 142092 Troitsk, Moscow r-n, Russia \\
 $^4$School of Physics and Space Research, University of Birmingham,
Birmingham~B15~2TT, United Kingdom\\
}
\date{November 22, 1993}
\maketitle
\def\v{\varepsilon}
\def\vv{\mbox{$|\v_i-\v_j|$}}\newlength{\len}
\def\ff{f(|x\!-\!x'|)}
\def\ip#1{\int_{|#1|\ge \frac{s}{2}}\!\!\!\!d#1\,}
\def\ix{\ip{x'}}
\def\ixx{\frac{1}{2}\ip{x}\ix }
\def\av#1{\langle#1\rangle}
\def\N{\av N}
\def\NN{\av{N^2}-\N^2}
\def\rta{\rightarrow}
\def\rti{\rta\infty}

\begin{abstract}
For a disordered system near the Anderson transition we show that the
nearest-level-spacing distribution has the asymptotics $P(s)\propto \exp(-A
s^{2-\gamma })$ for $s\gg \av{s}\equiv 1$ which is universal and intermediate
between the Gaussian asymptotics in a metal and the Poisson in an insulator.
(Here the critical exponent $0<\gamma<1$ and the numerical coefficient $A$
depend only on the dimensionality $d>2$).
It is obtained by mapping the energy level distribution to the Gibbs
distribution for a classical one-dimensional gas with a pairwise interaction.
The interaction, consistent with the universal asymptotics of the two-level
correlation function found previously, is proved to be the power-law repulsion
with the exponent $-\gamma$.
\end{abstract}

\pacs{}

\rightmargin\leftmargin\advance\leftmargin -3cm
It is well known \cite{GE,Ef:83,AS} that a statistical description of energy
levels of quantum disordered systems in the metallic phase is provided by the
random matrix theory (RMT)\cite{RMT}. Its most important characteristic is the
repulsion between the energy levels at any scale. In the Anderson insulator
phase, the energy levels are independent and described by the Poisson
statistics, provided that the appropriate states are separated by the length
exceeding the localization length.

It has been conjectured\cite{AS2,ShSh} that a universal statistical
description is possible also in the critical region in the vicinity of the
Anderson transition between the two phases. The dimensional scaling estimation
made in Ref.\cite{AS2} for the variance $\NN$ of the number of energy levels
in a given energy interval, has suggested that it is proportional to $\N$,
thus being different from the Poisson statistics only by a certain number. A
different statistical characteristic, the nearest-level-spacing distribution,
has been conjectured in Ref.\cite{ShSh} on the basis of numerical simulations
to be some universal `hybrid' of the Poisson distribution for large level
spacings and the Wigner surmise (see below) for small spacing.

However, it has recently been analytically proved \cite{KLAA} that the
universal statistics, exactly applicable near the Anderson transition point
(mobility edge), is entirely new and drastically different from both the RMT
and the Poisson limit. The variance of the number of levels in the energy
interval $E\gg \Delta$ (centered at the Fermi energy) has been found as
\begin{equation}
\label{1}
\NN= \frac{a_{d}}{\beta}\left(\frac{E}{\Delta} \right)^{\!\gamma }
\equiv\frac{a_{d}}{\beta}\N^{\gamma }, \qquad
0<\gamma<1.
\end{equation}
Here $\Delta$ is the mean level spacing, $\av{\ldots}$ denotes the ensemble
averaging, the coefficient $a_{d}$ and the critical exponent $\gamma$
depend only on the dimensionality
$d>2$, and $\beta$ is determined by the class of symmetry ($\beta=1,2,$ or $
4$ for unitary, orthogonal, and symplectic ensembles, respectively
\cite{RMT}). Eq.\ (1) is exact at the mobility edge in the limit
\begin{eqnarray}
\label{lm}
L\rti,\qquad\N={\rm const}\gg1.
\end{eqnarray}
In the same limit, the metallic phase is exactly described by the RMT and the
insulator phase by the Poisson statistics.

Thus, the spectral rigidity does not disappear at the mobility edge (in
contrast to the insulating phase where the levels are independent and the
variance equals $\N$), but it is considerably weaker than in the metallic
phase (where the fluctuations are suppressed and the variance is proportional
to $\ln\N$). Then it is naturally to expect that the nearest-level-spacing
distribution near the Anderson transition is also different from those both in
the metal and insulating phases.

Indeed, we will show in this Letter that the asymptotics of this distribution
at the mobility edge is given by
\begin{eqnarray}
\label{ME}
P(s)\propto\exp(-A_d\beta s^{2-\gamma }),\qquad s\equiv\omega/\Delta \gg 1,
\end{eqnarray}
where $\omega$ is a distance between adjacent levels, and $A_d$ is some
numerical factor depending only on the dimensionality $d$. This is drastically
different from both the Poisson distribution, $P(s)=\exp(-s)$, and the exact
Gaussian asymptotics in the metallic phase \cite{RMT}
\begin{eqnarray}
\label{WS}
P(s) \sim \exp\left(-\mbox{$\frac{1}{16}$}\pi^2\beta s^2\right)
\end{eqnarray}
Note that the famous Wigner surmise $P(s)=(\pi s/2) \exp(-\pi s^2/4)$ (for
$\beta=1$) describes this asymptotics only approximately \cite{RMT}.

Both the universal variance (1), and the asymptotics of the distribution (3)
result from the exact asymptotics of the spectral density correlation function
at the mobility edge
\begin{equation}
\label{Ro}
R(\omega)\equiv \frac{1}{\nu_{0}^{2}} \Bigl<\nu(\v)\nu (\v')
\Bigr>-1=-{c_{d}}{\beta}^{-1}|x-x'|^{-2+\gamma}, \qquad x\equiv \v/\Delta,
\qquad |x-x'|\gg 1
\end{equation}
where $\nu(\v)$ is the exact density of states at the energy $\v$, $\nu_{0}$
is the mean density of states, $c_{d}$ is a positive number depending only on
the dimensionality $d>2$. The asymptotics (\ref{Ro}) has been obtained in
Ref.\cite{KLAA} by calculating all the diagrams (with accuracy up to a
numerical coefficient) which turned out to be possible after taking into
account the analytical properties of the diffusion propagator and certain
scaling relations at the mobility edge.

To derive the announced result (\ref{ME}), we will use the effective ``plasma
model'' as suggested by Dyson\cite{Dsn}. In such a model, the level
distribution is mapped to the Gibbs distribution of a classical
one-dimensional gas of fictitious ``particles'' with a repulsive pairwise
interaction $f(\vv)$ in the presence of a confining potential $V(\v)$
\begin{eqnarray}
\label{G}
{\cal P}(\{\v_n\}) =Z^{-1}\exp\left[-\beta{\cal H}(\{\v_n\}) \right],\\
\label{H}
{\cal H}(\{\v_n\})= \sum_{i<j}f(\vv)+ \sum_{i}V(\v_i)
\end{eqnarray}
Here $Z$ is the partition function and $\beta$ plays a role of the inverse
temperature. For $f(\vv) \equiv\ln\vv^{-1} $, Eqs.\ (\ref{G}), (\ref{H})
reproduce exactly the level distribution in the RMT, with $\beta$ depending on
the symmetry class as described after Eq.\ (1). The choice $V(\v_i)= \v_i^2$
for the confinement potential leads to the Gaussian ensembles but other
choices are also possible \cite{RMT,Bee}. In the metallic phase such a
description is exact for the energy separation
\mbox{$\vv<E_c\equiv\hbar/\tau_D $} where $\tau_D= L^2/D$ is the `ergodic'
time necessary for an electron to diffuse across the system. For $t<\tau_D$,
i.e.\ \mbox{$\vv>E_c$}, the level statistics is completely different \cite{AS}
from that of the RMT. However, it is also described with the Gibbs
distribution (\ref{G}), (\ref{H}), albeit with the pairwise interaction $f$
having the power-law asymptotics \cite{JPB}.

At the mobility edge $E_c\sim\Delta$ ($g=E_c/\Delta$ is a dimensionless
conductance). Therefore, the energy separation of a few $\Delta$ is already
outside of the RMT region of validity. We will show that the asymptotics of
the correlation function (\ref{Ro}) at the mobility edge is described
correctly by the Gibbs model with the power-law interaction
\begin{equation}
\label{pl}
\ff=\frac{1-\gamma}{2\pi c_d}\cot(\pi\gamma/2)\,|x-x'|^{-\gamma},\qquad
0<\gamma<1,\qquad x\equiv\v/\Delta.
\end{equation}
(Naturally, this interaction is different from that in Ref.\cite{JPB}, where
$\gamma>1$, which describes the {\em nonuniversal} level statistics in the
metallic phase at the scale $\omega\gg E_c$.) Before proving this, we will
show how the form of the pairwise interaction governs the asymptotics of
$P(s)$.

The distribution $P(s)$ describes the probability to find the nearest adjacent
level at the distance $s=\omega/\Delta$ from a given one. It is equivalent to
the probability to find a ``gap'' of the width $s$ (i.e.\ region that contains
no ``particles'') in the Gibbs model. This probability is obtained \cite{RMT}
from Eq.\ (\ref{G}) as
\begin{eqnarray}
\label{Ps}
P(s)=\exp\left[-\beta(F_s-F_0)\right]
\end{eqnarray}
where $F_s$ is the free energy of the one-dimensional gas (\ref{H})
distributed along the straight line with the gap $s$ around its center. For $
s\gg1$, one introduces a continuous density $\rho_{s}(x)$ to describe such a
distribution. Then, in the mean-field approximation (MFA) $F_s$ may be
expressed as the functional
\begin{eqnarray}
\label{Fs}
F_s=\ixx\rho_s(x)\rho_s(x')\ff +\ip{x}\rho_s(x)V(x),
\end{eqnarray}
where  $\rho_{s}(x)$ obeys the mean-field (MF) equation
\begin{eqnarray}
\label{MF}
\ix\rho_s(x')\ff=-V(x)-\mu_s,\qquad |x|\ge s/2.
\end{eqnarray}
Here $\mu_s$ arises from the ``particles'' number conservation (corresponding
to the level number conservation in the original quantum disordered system),
\begin{eqnarray}
\label{N}
\ip{x}{\rho_s(x)}dx=\int_{-\infty}^\infty {\rho_0(x)}= {\cal N}.
\end{eqnarray}
Taking $s=0$ in Eqs.\ (\ref{Fs}) and (\ref{MF}), one finds the density
$\rho_0(x)$ and free energy $F_0$ for a homogeneous distribution.

Equations similar to (\ref{Fs}) and (\ref{MF}) have been derived for a
circular ensemble (a classical gas with the $\log|x-x'|^{-1}$ interaction
confined to a circular wire) by Dyson \cite{Dsn} (see also Ref.\cite{RMT}) who
has also found the corrections to the MF solution (allowing for the entropy
term added to the functional (\ref{Fs}) and for the discreteness of the
original distribution) lead to a linear in $s$ contribution to the difference
$F_{s}-F_{0}$. For $s\gg 1$ this contribution is small compared to the leading
quadratic term, Eq.\ (\ref{WS}). We will consider only the terms leading in
the $s\gg 1$ limit which are described by the MFA.

For the circular ensemble \cite{Dsn} there was no need in the confining
potential $V(x)$. We use the linear ensemble that is more convenient for
relating the interaction $f(\vv)$ to the correlation function (\ref{Ro}).
Equation (\ref{MF}) with the weakly singular kernel (\ref{pl}) can be solved
for any $V(x)$. This exact solution shows a strong dependence of both $\rho_s$
and $\rho_0$ (and thus $F_s$ and $F_0$) on $V(x)$ (see Eq.\ (\ref{Sol}) for
$\rho_0$). However, it is easy to show that the difference $F_s-F_0$, and thus
the distribution (\ref{Ps}), does not depend on $V$ for $s\gg 1$ in the limit
${\cal N}\rti$. Furthermore, the asymptotics of this universal distribution
may be found, with accuracy up to a numerical coefficient, without knowing the
explicit solution to Eq.\ (\ref{MF}).

The explicit dependence of $F_s-F_0$ on $V(x)$, Eq.\ (\ref{Fs}), is excluded
straightforwardly with the help of Eqs.\ (\ref{MF}), (\ref{N}). Then, after
some transformations using the fact that the change in the ``chemical
potential'' due to the gap formation $\mu_s-\mu_0\sim s/{\cal N}\ll 1$,
one finds with the accuracy up to $s/{\cal N}$ that
\begin{eqnarray}
\label{Fsd} {\lefteqn {F_{s}-F_{0}=}}\nonumber\\
&&-\ixx \delta\rho_{x}
\delta\rho(x')\ff +\frac{1}{2}\int_{-s/2}^{s/2}dx\int_{-s/2}^{s/2}dx'
\rho_{0}(x)\rho_{0}(x')\ff
\end{eqnarray}
where $\delta\rho(x)=\rho_{s}(x)-\rho_{0}(x)$ decreases rapidly for $x\gg s$.
The function $\delta\rho(x)$ obeys the MF equation:
\begin{equation}
\label{IE} \ix\delta\rho(x')\,\ff=
\int_{-s/2}^{s/2}dx'\rho_{0}(x')\,\ff, \qquad |x|\ge s/2,
\end{equation}
which follows from Eq.\ (\ref{MF}), if one neglects the small term
$\mu_s-\mu_0\sim s/{\cal N}$. The homogeneous level density $\rho_{0}(x)$ in
Eqs.\ (\ref{Fsd}), (\ref{IE}) still depends on $V(x)$, Eq.\ (\ref{MF}).
However, for $|x|<s/2\ll {\cal N}$, this dependence is negligible, and
in the limit ${\cal N}\rti$ one finds $\rho_{0}=1$ (in units of $1/\Delta$).
Now it is clearly seen from Eqs.\ (\ref{Fsd}) and (\ref{IE}) that the quantity
$F_{s}-F_{0}=-\beta^{-1}\ln P(s)$ is determined by the interaction $\ff$ only.

Equations (\ref{Fsd}) and (\ref{IE}) are valid for an arbitrary long-ranged
interaction $\ff$. For the case of the power-law interaction (\ref{pl}) (with
$0<\gamma<1$), one may rescale $x\rta sx$ and $x'\rta sx'$ to find that the
solution of Eq.\ (\ref{IE}) (with $\rho_{0}=$const) has the form $
\delta\rho(x)=\varphi(x/s)$, where $\varphi(x)$ is a universal function
independent of $s$. Substituting this solution to Eq.\ (\ref{Fsd}) we arrive
after the same rescaling at: \begin{equation} \label{PL}
F_{s}-F_{0}=-\beta^{-1}\ln P(s)=A_{\gamma}\,s^{2-
\gamma} \end{equation} where the
universal constant $A_{\gamma}$ depends only on the power of the interaction.
Calculating this constant for the limiting case ($\gamma=0$) of the
logarithmic interaction in Eqs.\ (\ref{Fsd}), (\ref{IE}) one reproduces the
asymptotics (\ref{WS}) known from the RMT \cite{RMT} which leads to the
announced result Eq.\ (\ref{ME}).

To prove that the interaction (\ref{pl}) reproduces the correlation function
(\ref{Ro}), we use the relationship \cite{Bee}:
\begin{equation}
\label{VD} R(x,x')=-\beta^{-1}\frac{\delta \rho_{0}(x)}{\delta V(x')}.
\end{equation}
For the interaction $f=a_{\gamma}|x-x'|^{-\gamma}$ and an arbitrary
$v(x)\equiv
V(x)+\mu_0$, the solution to Eq.\ (\ref{MF}) with $s=0$ is found, using the
methods described in Ref.\cite{IE}, as follows:
\begin{eqnarray}
\label{Sol}
\rho_0(x)&=&\frac{\cos^{2}(\pi\gamma/2)(x+D)^{\frac{\gamma-1}{2}}}{\pi^{2}
a_{\gamma}}B(\gamma,\mbox{$\frac{1-\gamma}{2}$})\times\nonumber \\ &\times &
\frac{d}{dx}\left\{
\int_{x}^{D}dt\,(t+D)^{1-\gamma}(t-x)^{\frac{\gamma-1}{2}}\:\frac{d}{dt}
\int_{-D}^{t}d\tau\,(\tau+D)^{\frac{\gamma-1}{2}}(t-\tau)^{\frac{\gamma-1}{2}}
\;v(\tau)\right\},
\end{eqnarray}
Here $B$ is the Euler function, $D$ is the band edge that may be found from
Eq.\ (\ref{N}) and tends to infinity when ${\cal N}\rti$. Taking the
variational derivative (\ref{VD}), i.e.\ substituting $-\beta^{-1}
\delta(\tau-x')$ for $v(\tau)$ in Eq.\ (\ref{Sol}), one finds in the limit
$D\rti$:
\begin{equation}
\label{RR} R(x,x')=-\beta^{-1}\frac{1-\gamma}{2\pi
a_{\gamma}}\cot\left(\frac{\pi\gamma}{2} \right)\,|x-x'|^{\gamma-2}.
\end{equation}
So the Gibbs model with the power-law interaction results in the asymptotics
(\ref{Ro}) of the correlation function. Comparing Eqs.\ (\ref{Ro}) and
(\ref{RR}), we obtain Eq.\ (\ref{pl}).

Note that for all the three universal statistics, in the metal and insulating
phases and at the mobility edge, a simple relation holds between the variance
of the level number fluctuations in the limit (\ref{lm}), and the asymptotics
of the nearest-level-spacing distribution. Namely, if the variance
proportional to $\N^\gamma$, then $-\ln P(s)\propto s^{2-\gamma}$. The linear
in $\N$ variance is forbidden at the mobility edge \cite{KLAA} by the exact
sum rule that is due to the conservation of the total number of states ${\cal
N}$. Therefore, the Poisson (i.e.\ linear in $s$) asymptotics of $P(s)$ is
equally forbidden at the mobility edge. Finally, following Ref.\ \cite{ShSh},
we note that for $s\ll 1$ the distribution $P(s)$ shows at the mobility edge
the same behaviour as in the metallic phase, $P(s)\sim s^\beta$, which follows
from the general symmetry theorem proved by Dyson \cite{Dsn}. Then, the whole
distribution could be described by the following surmise:
 \begin{eqnarray}
 P(s)=B s^\beta \exp\left(-A_d \beta s^{2-\gamma}\right)
 \end{eqnarray}
where $B$ is found from the normalization conditions.

\end{document}